\documentclass{article}

\pdfoutput=1

\usepackage[style=numeric,maxnames=100 , backend=biber, citestyle=numeric]{biblatex}

\usepackage[english]{babel}
\usepackage[utf8]{inputenc}
\usepackage{amsmath}
\usepackage{graphicx}
\usepackage{subcaption} 
\usepackage[colorinlistoftodos]{todonotes}
\usepackage{booktabs}
\usepackage{csquotes}
\usepackage{hyperref}

\DeclareUnicodeCharacter{2212}{-}

\bibliography{biblio, biblio_2}

\title{Neural Sampling by Irregular Gating Inhibition of Spiking Neurons and Attractor Networks}

\author{Lorenz K. M\"uller, Giacomo Indiveri}

\date{}

\begin{document}
\maketitle

\begin{abstract}
A long tradition in theoretical neuroscience casts sensory processing in the brain as the process of inferring the maximally consistent interpretations of imperfect
sensory input.  Recently it has been shown that Gamma-band inhibition can enable neural attractor networks to approximately carry out such a sampling mechanism. In this paper we propose a novel neural network model based on irregular gating inhibition, show analytically how it implements a Monte-Carlo Markov Chain (MCMC) sampler, and describe how it can be used to model networks of both neural attractors as well as of single spiking neurons. Finally we show how this model applied to spiking neurons gives rise to a new putative mechanism that could be used to implement stochastic synaptic weights in biological neural networks and in neuromorphic hardware. %that requires a cheap source of stochasticity for plasticity or inference.

\end{abstract}

\section{Introduction}
%\paragraph{Purpose of this Paper}
%In this paper we present a mathematical model of neural dynamics that can express an MCMC sampler.  This mathematical model is an analytically tractable reformulation of a biologically motivated model of Winner-Take-All (or attractor) networks and cortical gamma-oscillations \cite{Mostafa_etal15} we previously proposed. Further we show that the same mathematical model can be interpreted as describing single spiking neurons under intermittent shunting inhibition. 
A fundamental question in computational neuroscience is related to how the brain can reconcile noisy sensory inputs and its internal models of the world to reach maximally consistent abstract interpretations of the physical causes of the sensory input.  It has been argued that the dynamics and spiking behavior of neurons can be interpreted as an expression of a sampling process by means of which the brain performs such inference operations \cite{fiser2010,Berkes_etal11}. This approach is appealing because it allows drawing parallels between neural dynamics and well-established algorithms such as Markov-Chain Monte-Carlo (MCMC) sampling \cite{Buesing_etal11,Petrovici_etal13,Metropolis_etal53} or restricted Boltzmann-Machines (RBM) \cite{Merolla_etal10,Neftci_etal13,Smolensky86}, and various constraint satisfaction problem (CSP) solvers \cite{Habenschuss_etal13,Mostafa_etal15b}.

A key problem in such models is related to how probabilistic behaviour can be implemented within these models. 
% noise is generated; or differently put, what physical mechanism allows neurons to behave stochastically.
Up to now there have been two major approaches proposed: either the neuron model has been assumed to be the intrinsically stochastic \cite{Buesing_etal11} (which is non-ideal for corresponding hardware implementations and incomplete from a theoretical point of view); or given a deterministic neuron model, such as an integrate-and-fire one, high-rate, weak Poisson input, that stochastically keeps the neurons close to the firing threshold has been used  \cite{Petrovici_etal13}.   
%In this paper we focus on a different approach that is based on the interactions between coupled oscillators that have slightly different frequencies (e.g., arising from device mismatch in hardware implementations). In  \cite{Mostafa_etal15}, we have already shown how networks of neurons subject to  gating inhibitory signals, e.g., derived from local sources of gamma oscillations, can produce stochastic search \ldots 

We focus on a third approach in which stochasticity arises from a sparsely relieved gating inhibition that can be thought of as a model of gamma oscillations. This is a computationally efficient approach as the required source of randomness is one Poisson spiketrain of fixed rate per neuron and even simpler sources of randomness, such as mismatched oscillations can suffice in practice \cite{Mostafa_etal15, Mostafa_Indiveri15}.

A by-product of this approach is a new mechanism that could underlie the implementation of stochastic synaptic weights and weight updates, both in biology and in neuromorphic systems. Stochastic synaptic weights (and updates) are of interest because they may improve learning capabilities \cite{Srivastava_etal14, Neftci_etal15} and allow for novel models of neuronal sampling, such as the one we suggest in this paper.
%These models study intrinsically stochastic neurons whereas we focus on deterministic neuron and attractor dynamics under oscillatory \cite{Mostafa_etal15} and Poissonian gating inhibition (that models the effect of gamma-band inhibition).

%In related works we have applied similar distributed Markov chain samplers to potential VLSI implementation of artificial neural networks \cite{Mostafa_Indiveri15,LorenzThesis} and CSP solvers \cite{Mostafa_etal15b}.

\section{Theory and Results}
\subsection{Abstract Sampling Model}
\label{sec:PBM}
In this section we will describe an abstract, mathematically tractable model of biological neural networks (that we will refer to as the Gating Inhibition Sampling Network or GISnet). Nodes in this network change their states at Poisson distributed times that correspond to periods of lifted gating inhibition; communication between nodes is mediated by weights that are random variables and nodes take new states according to which of their possible values received maximal input. 

\begin{figure}[t]
\centering
\includegraphics[width=0.9\textwidth]{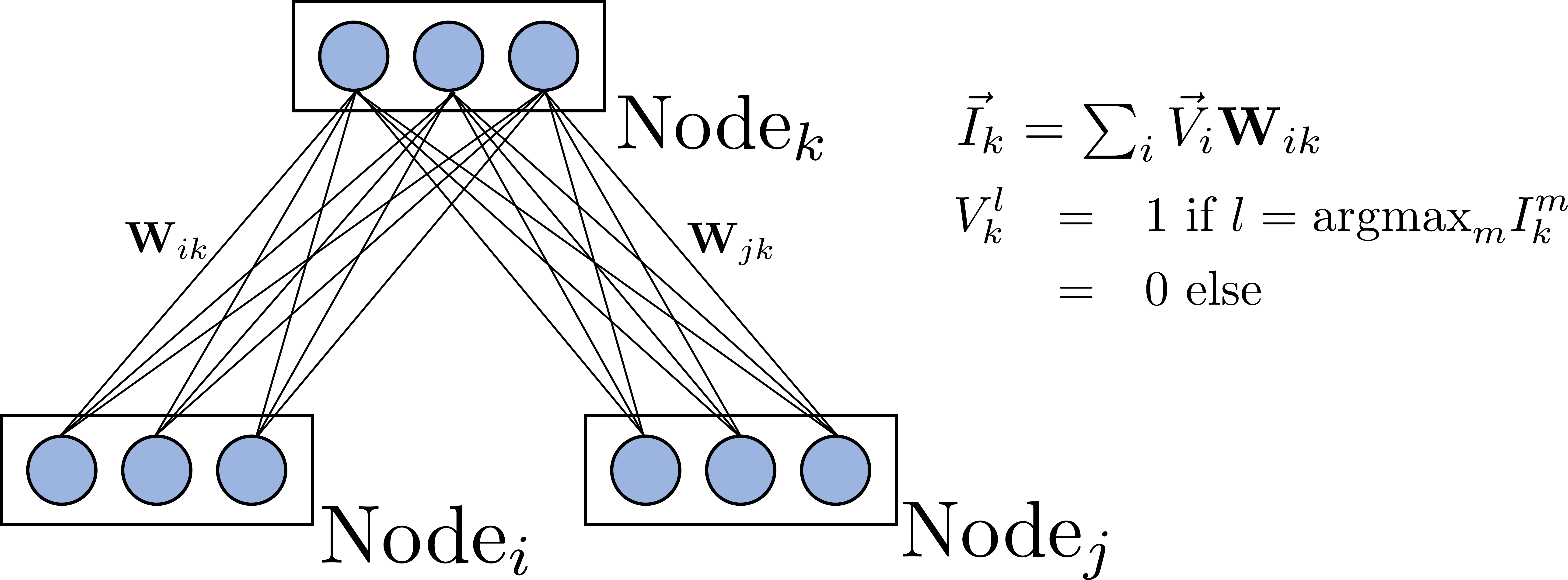}
\caption{A GISnet consisting of three nodes; the inputs and outputs to these nodes are vectorial. In $\vec{I_k}$ inputs from `active' presynaptic values are aggregated; at input independent `spike-windows' the node `activates' the value with the highest summed input. The weight matrices of the directed connection edges like $\mathbf{W}_{ik}$ are random variables.}
\label{fig:PBM}
\end{figure}

A GISnet, see figure \ref{fig:PBM}, consists of interconnected nodes that represent populations of neurons or single neurons. In the following superscripts will be used to denote the membership of a quantity to a node. Each node represents a variable. The state of node $i$ at any time is given by two vectors $\vec{I}_i(t)$, the input and  $\vec{V}_i(t)$ the output: $\vec{V}_i(t)$ represents the discrete output value of the variable corresponding to the node in a `one-hot encoding' (one entry is one, the others zero). If the $k$th entry of $\vec{V}_i(t)$, i.e. $V_i^k(t)$ is one, the corresponding variable has state $k$. The input vector $\vec{I}_i(t)$ keeps track of the weighted and summed inputs to the node. The vector $\vec{I}_i(t)$ has one entry $I_i^{k}$ for each possible value $V_i^k$ (i.e. the vectors $\vec{I}_i(t)$ and  $\vec{V}_i(t)$ have the same length).

Intuitively one may think of the vector $\vec{V}$ as the vector that identifies  which of $m$ mutually exclusive attractor states (or `patterns') receives the highest input in the input vector $\vec{I}$. This (and only this) `winning' population then may be able to influence the competitions in other nodes.

The nodes communicate through connections between different possible values; each connection between node $i$ and $k$ is weighted according to a weight matrix $\mathbf{W}_{ik}$ that specifies the impact state $V^q_i$ of node $i$ has on input channel $I^r_k$ of node $k$. Note that for each pair $i,k$ there is a full weight \emph{matrix} (corresponding to connections between different possible values nodes $i$ and $k$ can encode).

A GIS-node can only change its output value at special time \emph{points} that we will refer to as `spike windows' or `update times'. Let $s^i(t)$ be the function that takes value one, in spike windows of $i$ and zero otherwise. 
The spike-windows $S^i = \lbrace t | s^i(t) = 1 \rbrace$ of node $i$ are Poisson-distributed in time with a fixed average density of $R$ (the spike-rate of the node); notably these spike-windows are determined `a priori' in the sense that they are independent of the input node $i$ receives. 
%In contrast to these windows of opportunity for state changes, the times at which changes into a particular state occur, are \emph{not} necessarily Poisson distributed.
The `a priori' determined spike windows are a key difference to models like \cite{Buesing_etal11} or \cite{Nessler_etal09} and lend themselves to efficient neuromorphic implementation: Per neuron only a single Poisson spike train of similar rate as the desired output rate of that neuron is required, which is far more efficient than previously suggested methods in the vein of \cite{Petrovici_etal13}.

At a spike window node $i$ takes a new value $V^q_i(t)$ depending on its recent inputs. Specifically it takes the value $V^q_i(t)$ for which node $i$ received the highest weighted input, $q = \text{argmax}_x (I^x_i(t))$; the input $I^q_i(t)$ to node $i$ is given by
\begin{equation}
I^q_i(t) = \sum_{j} \vec{V}_j \mathbf{W}_{ij}
\label{eq:pbmInp}
\end{equation}
Where the sum over $j$ goes over all upstream connected neurons. In words, at a spike window, each possible value of the node sums its current input from connected variable values, weighted by the (random variable) weight matrices; then the value with the highest input becomes the node's new value. 

Note that it is irrelevant how often an upstream neuron has changed its state (`spiked') since the downstream neuron's last spike window, only the current state at the update time matters. If there is ambiguity about which value received the highest input (i.e. if there is no unique $\text{argmax}_x (I^x_i(t))$) the node maintains its previous value (other `tie-breaking' mechanisms could be used).%( `tie-breaking' mechanisms other than this `hysteretic' one could also be used.

Since the spike-times are independent of other state variables, we can analyse this continuous time system as a discrete time system: Each discrete time-step corresponds to a time $t_s$ at which one of the nodes may change its state. Then we can say for the discrete time system that at each round one (uniformly) random node undergoes a state change (this state `change' may be the trivial transition from a state to itself). 

Notably the corresponding discrete-time system is Markovian, which simplifies its analysis. 

\subsection{Single Neurons}
\label{sec:PBMsingle}
\begin{figure}
\centering
\begin{subfigure}[b]{0.45\textwidth}
\includegraphics[width=\textwidth]{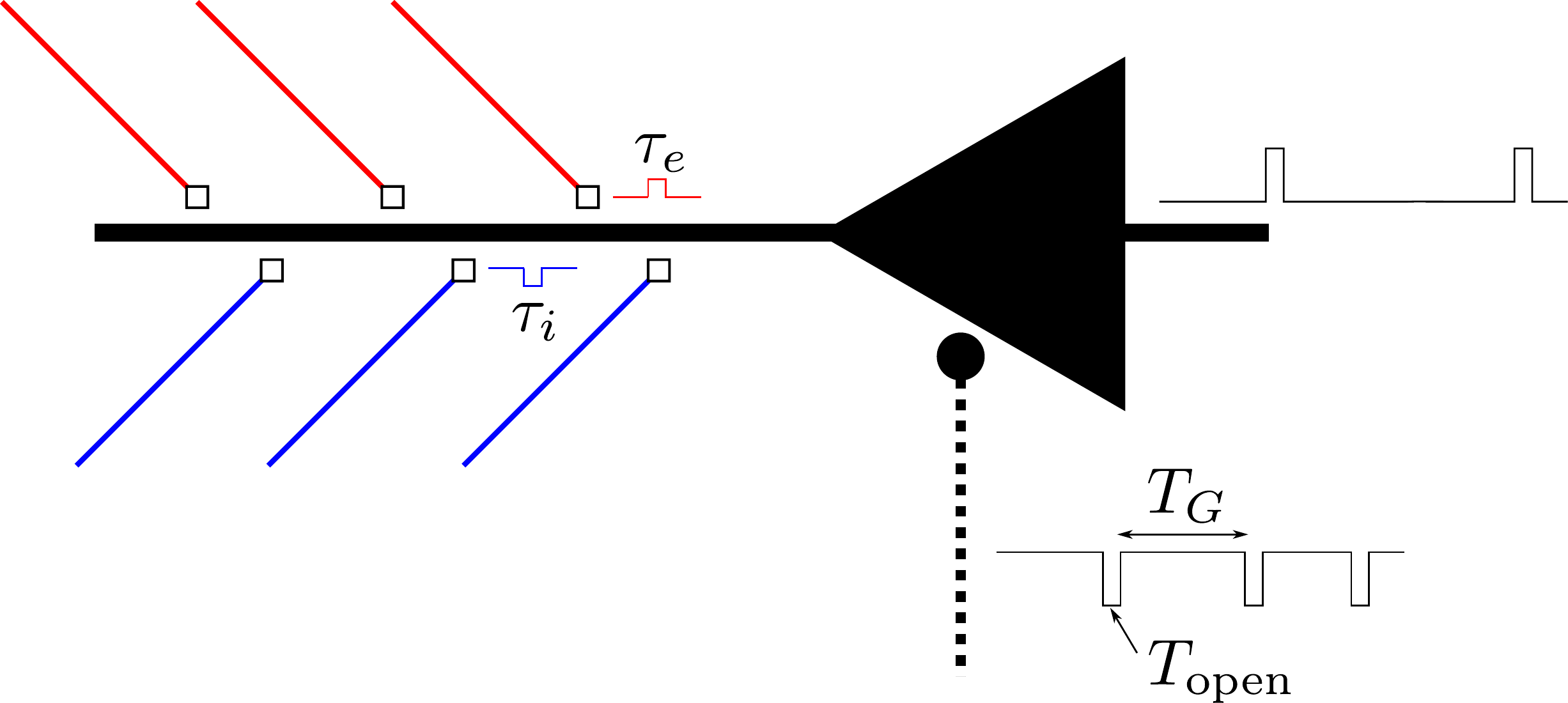}
\subcaption{Schematic neuron modelled by GIS node}
\label{subfig:PBMbio}
\end{subfigure}
\quad
\begin{subfigure}[b]{0.30\textwidth}
\includegraphics[width=\textwidth]{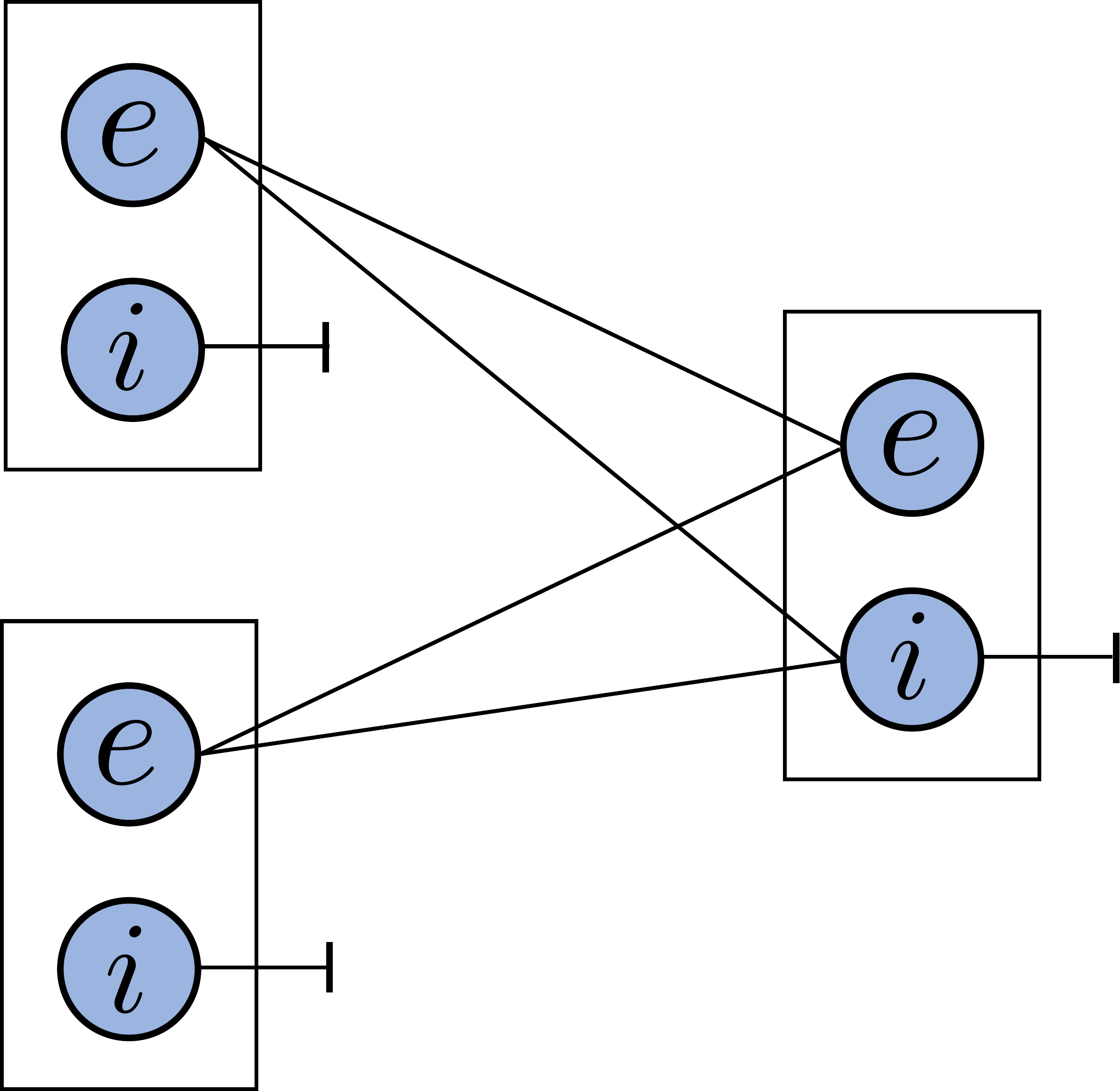}
\subcaption{A small GISnet with such nodes.}
\label{subfig:PBMtheo}
\end{subfigure}
\caption{(\subref{subfig:PBMbio}) A neuron under gating inhibition that releases at Poisson distributed times (rate $1/T_\text{G}$) for a short time ($T_\text{open}$). At these times it evaluates whether the sum of EPSPs (time constant $\tau_e$) or IPSPs (time constant $\tau_i$) is greater. Typically assume $\tau_i \approx \tau_e$, $T_\text{open} \ll \tau_e$ and $\tau_e < T_G$. (\subref{subfig:PBMtheo}) If the neuron gets above threshold input (i.e. more $e$xcitatory than $i$nhibitory input $I^e > I^i$) it emits a spike (i.e. goes to state $V^e$). To model a spiking neuron connection weights out of $V^i$ are zero, those out of $V^e$ are random variables that model the interplay of postsynaptic gating inhibition and PSP shape.}
\label{fig:singlePBM}
\end{figure}

Here we layout how spiking neurons can be modelled as GIS nodes. In figure \ref{fig:singlePBM} the GIS node used to implement a single neuron is shown.

Let us consider a neuron that receives strong shunting inhibition that is at Poisson times briefly lifted. During that short open window the membrane of the neuron will see the current influx through other activated ion channels and will evaluate, whether their combined conductance is high enough to overcome its spiking threshold; in other words, it measures whether it has received higher input to the inhibitory or excitatory `mode' (for a typical biological neuron with a bias to the inhibitory one). In the latter case, the neuron sends a spike to all downstream connected neurons that gets multiplied by the synaptic weight; this spike may or may not have an effect on the downstream neuron, depending on whether it is relieved of its gating inhibition within the post-synaptic potential (PSP) time constant ($\tau_e$ or $\tau_i$). This varying impact of spikes is modelled in the GISnet by the fact that weights are random variables. The case of inhibition exceeding excitation, in which no output (spike) is produced, can be incorporated into a GISnet simply by making all outgoing connections of the `inhibitory' node state have zero weight. 

Let us consider the distribution of weights we need to assume to correctly describe a spiking neuron as a GIS node. If we assume some fixed, post-synaptic kernel and a sufficiently short window during which the neuron's gating inhibition is released and it integrates its inputs on to the membrane potential the effective weight the incoming spike will have, is simply the integral over the post-synaptic kernel for the duration of the time window; to obtain a probabilistic weight we assume (as previously) that the presynaptic spike time and the release of inhibition occur at random time points. Note that in this way the effective weight distribution is generated by the interplay of the post-synaptic kernel shape and the time-constant of the lifted gating inhibition. For a gating window of duration $T_G$ and a gating function
\begin{equation}
G(t) = \begin{cases}
0 & \text{if $t<T_\text{open}$} \\
1 & \text{otherwise}
\end{cases}
\end{equation}
%the case of independent Poisson distributions for pre- and postsynaptic spike windows and with $G(t)$ the gating window, we obtain an effective synaptic weight of
we obtain
\begin{equation}
W_\text{eff} = \int_{0}^{\Delta t} \text{PSP}(t) \cdot (1-G(t + \delta t)) dt
\end{equation}
where $\text{PSP}(t)$ is the post-synaptic potential at time $t$, $\Delta t$ is its maximal duration and $\delta t$ is a random variable that expresses the time that passed between the arrival of the presynaptic spike and the onset of the relief of gating inhibition on the postsynaptic neuron. $\delta t$ is a random variable and thereby $W_\text{eff}$ is a random variable too. In the case of independent Poisson distributions for pre- and postsynaptic spike windows, $\delta t$ is distributed according to
\begin{equation}
p(\delta t) \propto \exp(t / \lambda)
\end{equation}
where $\lambda$ is the rate of the postsynaptic spike windows. Note that for a box-shaped postsynaptic potential and $\Delta t \ll T_G$ we obtain a Bernoulli distribution for $W_\text{eff}$.

If the timings of the pre- and post-synaptic spike windows can be controlled, e.g. through some coupling mechanism on the neurons providing the gating input, one could effectively control the weight distributions between the neurons. In other words, the network could thus dynamically change its connectivity to `load' a particular effective connectivity matrix by way of coupling the timing spike windows. In other words the GISnet allows naturally for an instantiation of the concept of `communication through coherence' \cite{Fries05}: The effective connectivity of the network can be modified by the coupling of the gating inhibition timings between various nodes.

%\begin{equation}
%p(w=x) = \frac{1}{Z} \int_0^\infty \delta(\hat{\text{PSP}}(\Delta t),x) \exp(\Delta t / \lambda_{out}) d\Delta t 
%\label{eq:weightDist}
%\end{equation}
%where $\hat{\text{PSP}}(\Delta t)$ is the postsynaptic potential after a time of $\Delta t$ has passed since the presynaptic spike integrated over $\tau_G$,
%\begin{equation}
%\hat{\text{PSP}}(\Delta t) = \int_{\Delta t}^{\Delta_t + \tau_G} PSP(t) dt,
%\end{equation}
%$\lambda_\text{post}$ is the rate of the postsynaptic spike windows and $Z$ is a normalization constant. To incorporate this into the PBM model one simply modifies Equation \ref{eq:pbmInp} to
%\begin{equation}
%I^i_k(t) = \sum_{j,l} v^j_l(t) \hat{W}_{i,j,k(t),l(t)} .
%\end{equation}
%where $\hat{W}$ is distributed according to Eq. \ref{eq:weightDist}. 

A simple biological observation complicates this picture somewhat and deviates from the GIS model: A neuron can spike more than once, if it receives a very high input.
This issue steps outside the proposed GIS model. It can be dismissed if we assume a neuron model whose refractory time constant is longer than the spike window opened by intermittently inactive shunting inhibition. However, the possibility of spiking more than once makes the network more `expressive' (there is a greater number of possible states) and could have useful effects. Namely if the single neuron has a linear transfer-function, this would lead to a linear response with Gaussian noise in the limit of a high fan-in, low leak neuron (see section \ref{subsec:singleNode}). 
%A linear transfer function with Gaussian noise is interesting as it has been used in generalized Boltzmann Machines to obtain models with continuous valued variables \cite{?}. We will address this situation in greater detail further on. 

\subsection{Attractor Networks}
\cite{Mostafa_etal15} studied WTA networks under oscillatory inhibition and demonstrated that they can be modelled by an MCMC sampler: The MCMC operator given in therein with purely empirical justification, is in fact the MCMC operator of a GISnet (detailed comparisons of the network model and the MCMC sampler are given there as well). Thus we provide here the theoretical top-down complement to the bottom-up approach of that paper. 

Notably irregular gating inhibition with periods of relief that are particular to single nodes can be interpreted as a model of gamma-oscillations \cite{Mostafa_etal15}. In summary this is the case, because the interneurons that mediate gamma oscillations inhibit their targets mostly perisomatically and because gamma-rhythms originate locally \cite{Buzsaki_Wang12}. Perisomatic inhibition is particularly effective in hindering its targets from firing (while less impacting the integration of currents that takes place on the dendrites) and the local origin of the gamma-rhythms implies that at the very least different local neighbourhoods of neurons, are driven with differing periods. 

In practice the local gamma periods are not perfectly stable, but the period lengths vary \cite{friedman2000dynamics}. In other words the true distribution of the `spike-windows' mentioned earlier lies somewhere between the analysable Poisson distribution we assumed here and the oscillatory case simulated in \cite{Mostafa_etal15} (notably the latter two produce very similar high level behaviour).

Intuitively the mapping between a GISnet and a network of irregularly inhibited WTA nodes can be understood as follows: Each node of a GISnet models one WTA unit and each possible value of the node corresponds to one excitatory population in the WTA. The random nature of the connections comes about by the relative timings of the irregular inhibitory signals to the various WTA nodes; if a node $x$ is completely unihibited while the inhibition on $y$ is lifted, it affects it with the full weight, otherwise with a down-scaled weight. These relative timings occur pseudo-randomly.
%We refer the reader to \cite{Mostafa_etal15} for further details.

\subsection{The MCMC Transition Matrix}
\label{sec:PBMandMCMC}
%In section \ref{sec:cts} we outlined a method to define or measure the probability distribution $p_{CTS}$ produced by any CTS. While this method makes sense as a definition, it is of relatively little practical use: To find out what probability distribution is encoded by e.g. a GISnet with a given connectivity structure, this method requires us to simulate its continuous time behaviour. 

Here we will describe a method of obtaining the limiting distribution of a GISnet (if it exists). The key question is how to construct the transition matrix $T$ of the corresponding MCMC sampler, since this fully determines its behaviour. Let the system be in state $s_i$; what is the probability that it will transition into state $s_j$? We study three cases for different relationships between $s_i$ and $s_j$.

\begin{enumerate}
\item \emph{$s_i$ and $s_j$ differ in more than one variable.} In this case the probability $T_{ij}$ is set to zero, because it is impossible for multiple variables to change their state at the same time in a GISnet (time is continuous and state changes are instantaneous, so that the probability of two occurring simultaneously vanishes).

\item \emph{$s_i$ and $s_j$ differ in exactly one variable.} In this case $T_{ij} = P(v_{up}) \cdot P(v_{new} = v(s_{j}) | s_{i}, v_{up})$, where $P(v_{up})$ is the probability that the variable $v$ in which $s_i$ and $s_j$ differ is the one that changes its state and $P(v_{new} = v(s_{j}) | s_{i}, v_{up})$ is the probability that $v$ takes the value it has in $s_j$ given that the system state at the change is $s_{i}$ and that $v_{up}$ is the variable that changes its state. We will shortly address how to construct these probabilities from the network connectivity.

\item \emph{$s_i$ and $s_j$ are the same state}. In this case we can construct $T_{ij}$ using the fact that the previous item implicitly defines the probability $p_{\neq} = P(s_i \neq s_j)$ by the relation $p_{\neq} = \sum_{j \neq i} T_{ij}$, so that $T_{ii} = 1 - p_{\neq}$ by virtue of the normalization of $\sum_{j} T_{ij}$.   
\end{enumerate}

The probability $P(v_{up})$ is simply set to $\frac{1}{n_N}$ where $n_N$ is the number of nodes in the GISnet: Each node updates at Poisson distributed time points and it is therefore equally likely that any node is the next one to update (the update probability density is constant). 

For notational simplicity we will consider the special case where each variable has two possible states $\lbrace 0,\alpha \rbrace$ and all weights are Bernoulli variables of absolute value $\alpha$ or always zero. Then $\mathbf{W}_{ij,\text{eff}} = \mathbf{\hat{W}}_{ij} \cdot \mathbf{b}_{ij,p}$, where $\mathbf{b}_{ij,p}$ is a matrix of Bernoulli variables with probability $p$ and $\mathbf{\hat{W}}$ is a matrix whose entries are either zero or $\alpha$. The general case without these restrictions is straight forward to write down based on this simpler case, but is notationally cumbersome. Instead of summing up weights, in this case we simply count numbers of inputs to a particular state. Let $i$ be the updating node and let $u_i$ be the number of potential inputs to state 0 of $i$ and $t_i$ the total number of potential inputs to node $i$. The number of potential inputs to state 1 is then $t_i - u_i$. In formulas we define
\begin{equation}
t_i = \sum_{j} \vec{V}_j \mathbf{W}_{ij},
\label{eq:t(n)}
\end{equation} 

\begin{equation}
u_i = \sum_{j} V_{j}^{0} \vec{W}_{ij}^{0}.
\label{eq:u(n)}
\end{equation} 

By assumption every potential input is set to zero with probability $p$. In this formulation the node receives an effective number $t_{i,\text{eff}}$ of inputs. This number is distributed according to
\begin{equation}
P(t_{i,\text{eff}} = k) = \binom{t_i}{k} (1-p)^k p^{t_i-k}.
\end{equation} 

The aforementioned conditional transition probability $P(v_{new} = 0 | s_\text{prev}, v_{up})$ is the probability that most of the received inputs go to state 0:
\begin{equation}
P(v_{new} = 0 | s_\text{prev}, v_{up}) = \sum_{k=0}^{t_i} \underbrace{ \binom{t_i}{k} (1-p)^k p^{t_i-k}}_{P(t_{i,\text{eff}} = k)} \underbrace{\sum_{l=0}^{k/2} \frac{\binom{u_i}{k-l}\binom{t_i-u_i}{l}}{\binom{t_i}{k}}}_{\text{$P(u_i>t_i-u_i)$}},
\end{equation}
which simplifies to
\begin{equation}
P(v_{new} = 0 | s_\text{prev}, v_{up})  = \sum_{k=0}^{t_i} (1-p)^k p^{t_i-k}\sum_{l=0}^{k/2} \binom{u_i}{k-l}\binom{t_i-u_i}{l}
\end{equation}
 and finally yields
\begin{equation}
\label{eq:transition}
T_{ij} = \frac{1}{n_N} \sum_{k=0}^{t_i} (1-p)^k p^{t_i-k}\sum_{l=0}^{k/2} \binom{u_i}{k-l}\binom{t_i-u_i}{l}.
\end{equation}
Note that the above is dependent on the previous state and the target state because $u_i$ and $t_i$ depend on them.

Using the above construction, we can evaluate the probability distribution induced by a certain connectivity by finding the limiting distribution of the associated MCMC sampler. To get this one still has to solve a large system of linear equations (the limiting distribution is the eigenvector with eigenvalue one of the transition matrix). 

The construction of the limiting probability distribution would be simpler, if the transition matrix fulfilled `detailed balance' ($T_{ij} = T_{ji} $). This is however not the case in general for a GISnet, similar to the neuronal MCMC sampler suggested in \cite{Buesing_etal11}.

%This is somewhat dissatisfying as the problem of how to construct a weight matrix that produces a desired distribution (or models a set of given samples) is still open (because it is difficult to invert the process that lead from connectivity to limiting distribution). In the next section we will tackle this problem from a different avenue: the single node perspective.

\subsection{The Single Node Activation Function}
\label{subsec:singleNode}
A more concise way of describing nodes in a GISnet can be obtained by studying their individual activation functions. The activation function expresses what the output response of a single node is for a given network state and connectivity. 

By definition we need to evaluate which of the competing states receives the highest input. For a binary node this means we want to find (for given network state $\vec{v}$, and weight tensor $W$) the probability that the input to one node state exceeds the input to the other. We can equivalently asses the probability that the difference of the two inputs exceeds zero. In this formulation it becomes clear that a binary GIS node is equivalent to a McCulloch-Pitts-Neuron \cite{McCulloch_Pitts43} with stochastic synapses.. Consider the activation $a$ of a McCulloch-Pitts-Neuron with Bernoulli input weights:
\begin{equation}
a_i = I^{1}_i - I^{0}_i = \sum_{j} V_{j}^1 (\vec{W}^1_{ij} - \vec{W}^0_{ij})  =  \sum_{j} \vec{V}_j \mathbf{U}_{ij}
\end{equation}
where $V^j$ are the presynaptic activities, $\mathbf{U}_{ij}$ is a redefined connection weight (the difference between the weights to the two states)
%and $b_p$ is a Bernoulli variable with probability $p_\text{on}$ (note the absence of the sum over $l$ compared to Eq. \ref{eq:pbmInp}, this falls away because the node only has a single state that is compared against the threshold zero in the McCulloch-Pitts formulation). 
Further we assume $\vec{V}^j \mathbf{U}_{ji} = q_j \cdot b_p(i,j)$ to be fixed up to a Bernoulli variable $b_p(i,j)$ with probability $p_{on}$; then we can write
\begin{equation}
a_i = \sum_j q_j b_p(i,j).
\end{equation}
$a_i$ is now a sum over weighted Bernoulli trials. The central limit theorem states that the mean of a sufficiently large number of independent random variables is approximately normally distributed \cite{bohm10} (if the random variables have well defined means and variances, which is the case here). Formulaically this yields for a neuron with sufficiently many inputs
\begin{equation}
a_i = \sum_j q_j b_p (i,j) \approx \mathcal{N}\left( \sum_i p_\text{on} q_i, \sqrt{ \sum_i p_\text{on} (1- p_\text{on}) q^2_i } \right),
\end{equation}
where the angular brackets denote an average. The probability that $a_i$ is greater than zero then is
\begin{equation}
p\left(\mathcal{N}\left(\mu, \sigma \right) > 0 \right) = \frac{1}{2} \text{erfc}\left(\frac{-\mu}{\sqrt{2}\sigma}\right),
\label{eq:gaussian}
\end{equation}
where we introduced $\mu = \sum_i p_\text{on} q_i $ and $\sigma =  \sqrt{ \sum_i p_\text{on} (1- p_\text{on}) q^2_i} $.
Note that the McCulloch-Pitts Neuron with Bernoulli distributed synaptic weights has recently also been studied in \cite{Neftci_etal15}; the `synaptic sampling machine' presented therein is closely related to a GISnet with binary nodes and Bernoulli distributed synaptic weights.

We can now also revisit the question from section \ref{sec:PBMsingle}, what the output rate of a neuron that can fire multiple times in one `spike window' would be: This is simply $a_i$, indeed a linear response (linear in $\sum_i q_i $) with Gaussian noise, however the variance of the noise increases with increasing mean. 

For higher order nodes (with more than 2 competing modes) we can make similar considerations. Here the input to a ~single state of the node must exceed the input to all others for that node to become active. The probability of state $i$ becoming active is then\footnote{note the correction compared to the previous version}
\begin{eqnarray}
p(i \text{ is active}) &=& \int_{-\infty}^{\infty} p( I_i = \alpha) \prod_{i \neq j} p( I_j < \alpha) d\alpha \\
&=& \int_{-\infty}^{\infty} \mathcal{N}(\alpha; \mu_i, \sigma_i) \prod_{i \neq j} \frac{1}{2} \left[ 1 + \text{erf} \left( \frac{\alpha - \mu_j}{\sqrt{2} \sigma_j} \right) \right] d\alpha
\label{eq:erfc}
\end{eqnarray}
where $\mu_i$ and $\sigma_i$ are the mean and variance of the Gaussian approximation to the input to state $i$ as introduced for the binary node.

\section{Conclusion}
In this paper we described an abstract neural network whose functional behaviour it is to sample in a Markovian manner from a probability distribution defined by its weights and detail how it maps both onto networks of individual spiking neurons as well as populations thereof acting as attractors.

We showed that this model can be viewed as a generalization of synaptic sampling machines \cite{Neftci_etal15} or McCullough-Pitts-Neurons \cite{McCulloch_Pitts43} with probabilistic synapses and highlighted a new mechanism that could underlie the generation of probabilistic synaptic weights. While it has been hypothesised that biological substrates can directly tune the full distribution of the postsynaptic potentials evoked by single synapses \cite{Aitchison_Latham15} and the reliability of neurotransimitter release of a particular synapse can indeed be highly variable \cite{branco_Staras09}, such hypotheses are as of yet unsubstantiated.

At the same time the proposed model describes analytically a good approximation of networks of neural attractors under gating inhibition, that constitutes a model of gamma oscillations \cite{Mostafa_etal15}. 

Neuromorphic systems currently lack the capability of cheaply implementing stochastic synaptic strengths. The here proposed mechanism could be implemented in any asynchronous neuromorphic platform that offers parametric changes of the PSP shape, such as \cite{Qiao_etal15}, with the simple addition of a gating functionality of the kind we described.

The computational efficiency of the generation of stochasticity in the model we propose, lends itself to direct mapping to neuromorphic hardware: Per represented variable only a single Poisson spike train of similar rate as the desired output rate of that neuron is required, which is more efficient than previously suggested methods in the vein of \cite{Petrovici_etal13}.

Finally the analytically well-described GISnet model will allow us to formulate theoretically motivated learning rules for networks of spiking neurons and attractor networks under gating inhibition in future work.

\section*{Acknowledgements}
We would like to thank Hesham Mostafa for pointing out an error in the previous version of this manuscript. This work was supported by the Swiss national science foundation grant Nr. \texttt{CRSII2\textunderscore 160756}. 
%\newpage

\printbibliography

\end{document}